\documentclass[english]{article}
\usepackage{graphicx}
\usepackage{amssymb}

\makeatletter

 \newcommand{\lyxaddress}[1]{
   \par {\raggedright #1 
   \vspace{1.4em}
   \noindent\par}
 }

\usepackage{babel}
\makeatother
\begin{document}

\title{Tachyons and EPR correlations}

\author{Bruno Cocciaro%
\thanks{b.cocciaro@comeg.it%
}}

\maketitle

\lyxaddress{Via Del Bosco 27, 56025 Pontedera (PI) Italy}

\lyxaddress{Liceo Scientifico XXV Aprile, Via Milano 2, 56025 Pontedera (PI)
Italy}

\begin{abstract}
No causal paradoxes will occur if a preferred reference frame for
tachyons propagation is assumed, and results of Bell's inequality
experiments may be well explained without using any \textit{telepathyc}
effect. We can read G. Faraci's and others' results, Lettere al Nuovo
Cimento, 15, 607-611 (1974), as a first quantitive indication on the
tachyons preferred reference frame velocity with respect to the Earth,
as well as on the tachyons velocity in their preferred reference frame.
In order to experimentally prove this assumption's validity, Aspect-like
experiments should be slightly modified.
\end{abstract}

\section*{Introduction.}

In this letter, I will show how it could be possible to explain the
Aspect results \cite{f} (and similar experiments) without using any
\char`\"{}telepathyc\char`\"{} effect%
\footnote{The literature on the EPR paradox is very extensive. I found G. C.
Ghirardi (1997) \cite{i} like a very powerful book to enter this
subject. In a certain sense, my reading of that book is the origin
of this paper. To have a report on Aspect's experiments as well as
other experiments on the Bell's inequality you may see F. Selleri
(1999) \cite{g} and references there shown.%
}. This is done by assuming the existence of a preferred reference
frame in which tachyons propagate isotropically.

I will follow this logical scheme:

\emph{a} - To have causal paradoxes we need tachyons + relativity
principle.

\emph{b} - Imagining that tachyons have a preferred reference frame
in which they propagate isotropically%
\footnote{Substancially we assume that relativity principle is not suitable
for tachyons (so as for sound) nor for events related with tachyons.
Of course all the other known physics remains unchanged. We'll only
suppose that wave function collapses are related with tachyons. We
do not mean that no other effect related with tachyons exists, but
only that, if any other exists, it is not yet known.%
} (so as sound) the causal paradoxes disappear.

\emph{c} - I suppose that tachyons mediate wave function collapse.

I mean this:

we have a couple of entangled particles%
\footnote{No matter which kind of particle. I imagine that the statements hereby
exposed could be applied to any kind of entagled particles.%
} and detect one of it. At the detection a tachyon leaves from the
point where the detection took place and go to \char`\"{}communicate\char`\"{}
at the other particle the result.

\emph{d} - using such tachyons the two measurements, \emph{m1}) and
\emph{m2}), could be no more \char`\"{}space like\char`\"{}:

\emph{m1}) detection of the left particle;

\emph{m2}) detection of the right particle.

\emph{e} - they could be no more \char`\"{}space like\char`\"{} because

the tachyon left from right reaches the left particle before its detection

or

the tachyon left from left reaches the right particle before its detection.

\emph{f} - There are particular experimental situations in which what
described in \emph{e} is not true. Measurements are not correlated
when these particular experimental situations are satisfied ($\Delta_{m}<\Delta<\Delta_{M}$
in following notations). This happens because, in such experimental
situations, the two measurements are really \char`\"{}space like\char`\"{},
it is really impossible that one could be the cause of the other.

\emph{g} - Perhaps G. Faraci and others (1974) \cite{b} casually
reproduced a uncorrelating experimental situation in one of its five
measurements.

\vspace{3mm}

I'll discuss points \emph{a} and \emph{b} in the appendix I, points
\emph{c}-\emph{f} in the part 1 and \emph{g} in the part 2.\vspace{3mm}

\section{EPR correlations and their possible link with tachyons.}

Let's assume the existence of a preferred inertial reference frame
\emph{R'} in which tachyons propagate isotropically. We call $V_{t}$
the magnitude of the velocity at which tachyons propagate isotropically%
\footnote{I mean that, if \emph{s} is the distance between the points \emph{P}
and \emph{Q}, then \emph{}a tachyon which left the point \emph{P}
when the clock fixed in \emph{P} was signing the instant $t_{in}$,
will arrive in the point \emph{Q} when the clock fixed in \emph{Q}
will sign the instant $t_{in}+s/V_{t}$. The reference frame \emph{R'}
is preferred because this is true for all the couples of points \emph{(P},
\emph{Q),} placed at an \emph{s} distance, only if \emph{P} and \emph{Q}
are fixed points in \emph{R'} (clocks was synchronized by using standard
relation). This is enough to say that relativity principle is not
suitable for tachyons.%
} in \emph{R'}. I am assuming that \emph{R'} clocks was synchronized
by using standard relation, that is by using light beams%
\footnote{Or, equivalently, using clocks transport synchronization. {}``Transport
synchronization'', in my opinion, means simply this: a clock travels
from \emph{A} to \emph{B.} It leaves from \emph{A} when it is signing
the instant $\tau$ and the clock fixed in \emph{A} is signing the
instant $\bar{t}$. At the arrival in \emph{B} the travelling clock
signs the instant $\tau+\Delta\tau$. If the distance between \emph{A}
and \emph{B} is \emph{d} and if we want synchronize by standard relation,
then the clock fixed in \emph{B} will be set at the instant $\bar{t}+\Delta\tau\sqrt{1+\left(\frac{d}{c\Delta\tau}\right)^{2}}$.
The clock motion must be uniform, no matter on the interval time value
$\Delta\tau$ measured by the travelling clock (that is no matter
on its velocity: the transport must not be {}``slow''). Uniform
motion means that, $\forall\alpha\in\left(0,1\right)$, the clock
signs the instant $\tau+\alpha\Delta\tau$ when its distance from
\emph{A} is $\alpha d$.%
} and fixing at $\bar{t}+l/c$ the instant of the clock when it receives
the light beam departed, from a point at a \emph{l} distance, when
the clock there fixed was signing the $\bar{t}$ instant.

Let's now assume that {}``entangled'' couples of particles are furthermore
characterized by the fact that when a measurement is done on one of
them a tachyon leaves and {}``communicates'' to the other particle
that it must {}``correlate'' to the result of the already performed
measurement (that is we imagine that tachyons are {}``hidden variables'').

Let's call \emph{R} the laboratory inertial reference frame and let's
assume that \emph{R'} moves at $\beta$\emph{c} speed ($-1<\beta<1$),
with respect to \emph{R,} along the \emph{x} axis direction (\emph{c}
is the light speed%
\footnote{That is the numerical value of \emph{c} is given by the number of
back and forth consecutive travels performed by a light beam on a
journey whose lenght is $\left(1/2\right)u_{l}$, where $u_{l}$ is
the unit length, to say that an assumed unit time interval $u_{t}$
is spent.%
}, and we assume that also in \emph{R} we have synchronized clocks
by using standard relation). For simplicity we treat the question
as if it is one-dimensional, so we are only interested in the tachyon
speed towards the two directions on the \emph{x} axis. Called $V_{t}^{+}$
the tachyon velocity toward the positive \emph{x} direction, and $V_{t}^{-}$
the velocity in the opposite direction, we obtain by the speed composition
law%
\footnote{I would remember that this law is an easy consequence of Lorentz transformations,
that is it follows by assuming \emph{R} and \emph{R'} both inertial
reference frame and $-1<\beta<1$. No matter on $\beta_{t}$ value.%
}:

\begin{equation} V_{t}^{+}=\frac{\beta_{t}+\beta}{1+\beta\,\beta_{t}}\,c\hspace{4mm};\hspace{4mm} V_{t}^{-}=\frac{\beta_{t}-\beta}{1-\beta\,\beta_{t}}\,c\label{1}\end{equation}

where we set $\beta_{t}\equiv\frac{V_{t}}{c}$ (it will be $\beta_{t}>1$)%
\footnote{Noticing that $\left(V_{t}^{+}<-c\vee V_{t}^{+}>c\right)\wedge\left(V_{t}^{-}<-c\vee V_{t}^{-}>c\right)$
is true for each $-1<\beta<1$ we obtain that in every inertial reference
frame tachyons propagate {}``faster than light'' toward any direction,
that is, sending instantaneously a tachyon and a photon from \emph{A,}
the tachyon will always arrive in \emph{B} before the photon. So we
are not assuming this but we are obtaining it as a consequence of
the velocity composition law and the existence of a preferred frame
in which tachyons propagate isotropically at a certain $\beta_{t}>1$.
Substancially, if a preferred reference frame in which tachyons propagate
{}``faster than light'' exists, then they will propagate {}``faster
than light'' in every inertial reference frame.%
}.

It could be probably useful to briefly discuss the above (\ref{1}).
It can be immediately noticed that both $V_{t}^{+}$ and $V_{t}^{-}$
can assume negative values: $V_{t}^{+}$ assumes negative values when
$\beta<-\frac{1}{\beta_{t}}$ and $V_{t}^{-}$ when $\beta>\frac{1}{\beta_{t}}$.
We just noticed that $V_{t}^{+}$ ($V_{t}^{-}$) is meant to be the
tachyon speed propagation toward the positive (negative) \emph{x}
axes direction. A tachyon propagating at $V_{t}^{+}<0$ ($V_{t}^{-}<0$)
is \emph{not} a tachyon propagating toward negative (positive) direction,
it is simply a tachyon that, while propagating toward the positive
(negative) direction, will meet along its travel clocks signing lower
instants%
\footnote{This fact doesnt give any problem. We could imagine analogue facts
in every day life that are perfectly comprehensible. Earth clocks
could be conventionally synchronized in the following way: each clock
will be set at 6.00 when it receives the signal {}``Sun is rising''.
Once clocks are synchronized it could happen that travelling by plane
from New Zealand to Italy we meet clocks signing always lower instants.
For example at the departure New Zealand clocks could sign 8.00 and
at the arrival Italian clocks could sign 7.00.%
}. That is, once clocks are synchronized by standard relation, a tachyon
leaving from point \emph{A} (\emph{cause}) when the clock fixed in
\emph{A} signs the $t_{A}$ instant, will reach point \emph{B} (\emph{effect})
- that belongs, with respect to \emph{A}, toward the positive (negative)
\emph{x} axes direction - when the clock fixed in the point \emph{B}
signs the $t_{B}<t_{A}$ instant. There is no cause and effect exchange
on this.

We could ask now:

if velocity doesn't give us the particle propagation direction, how
could we decide the {}``correct'' (that is, if the sign of $t_{B}-t_{A}$
does not tell us what the cause and the effect are, which is the physical
entity we must ask to know that)?

The answer to this question is in my opinion among the main results
that we could obtain by the full comprehension of the conventional
character of synchronization (to have a report on the conventionality
of simultaneity, see R. Anderson, I. Vetharaniam, G. E. Stedman (1998)
\cite{a}). Both the time interval measured by two clocks fixed on
different points, and velocity (that is strictly joined with that
time interval), are conventional entities that could never give us
physical meaningful results, but if we take into account also the
chosen synchronization as well. By the simple observation that $t_{B}-t_{A}<0$
(or $V_{t}^{+}<0$) we could not obtain any physical meaningful result.

The entity that gives us the particle propagation direction is its
momentum $\overrightarrow{p}$ that is not conventional like every
tridimensional vector given by the spatial components of tetravectors%
\footnote{I call here tetravector the vector written by means of its controvariant
components.%
} (see Anderson and others (1998) \cite{a} pag 127 and following).
I think that, as well as any other known particle, also tachyons propagating
toward positive (negative) direction should have $\overrightarrow{p}$
directed toward the positive (negative) direction, whatever the sing
of $V_{t}^{+}$ ($V_{t}^{-}$) (that sign is related with the conventionally
chosen synchronization whilst $\overrightarrow{p}$ is decided by
the experiment, not by our choices). I will discuss this point in
more detail in appendix II.\vspace {3mm}

Let's now imagine that a couple of entangled particles is created,
in \emph{R}, in the point $x=\bar{x}$, at the instant $\bar{t}$,
and that two detectors are placed respectively in $x=-d$ and $x=d$
($-d<\bar{x}<d$). We call $v_{1}=\beta_{1}c$ ($0<\beta_{1}<1$)
the magnitude%
\footnote{We are assuming here that the entangled particles have the same velocity
magnitude. If this is not true then the argument hereby stated should
be sligthly modified.%
} of the particles speed in the reference frame \emph{R}.

We'll call

Lparticle (or left particle) the particle directed toward the detector
placed in $x=-d$;

Rparticle (or right particle) the particle directed toward the detector
placed in $x=d$;

Ltachyon (or left tachyon) the tachyon which left, from $x=-d$ when
the detection was there performed, toward the Rparticle;

Rtachyon (or right tachyon) the tachyon which left, from $x=d$ when
the detection was there performed, toward the Lparticle.

Ltachyon travels at velocity $V_{t}^{+}$, whilst Rtachyon travels
at velocity $V_{t}^{-}$.

Rparticle will be detected at the instant $t^{+}=\bar{t}+\frac{d-\bar{x}}{\beta_{1}c}$
whilst Lparticle will be detected at the instant%
\footnote{With this we obviously mean that the clock fixed in $x=\bar{x}$ (in
\emph{R}) is signing the instant $\bar{t}$ when, in that point, particles
are created, the clock fixed in $x=d$ (in \emph{R}) is signing the
instant $t^{+}$ when, in that point, Rparticle is detected, and the
clock fixed in $x=-d$ (in \emph{R}) is signing the instant $t^{-}$
when, in that point, Lparticle is detected.%
} $t^{-}=\bar{t}+\frac{d+\bar{x}}{\beta_{1}c}$.

To let (a) Rparticle reach the detector placed in $x=d$ before it
is reached by Ltachyon, it must be:\begin{eqnarray*}
t^{+}<t^{-}+\frac{2d}{V_{t}^{+}} & \Longleftrightarrow & \frac{\bar{x}}{d}>-\beta_{1}\frac{1+\beta_{t}\beta}{\beta_{t}+\beta}.\end{eqnarray*}

To let (b) Lparticle reach the detector placed in $x=-d$ before it
is reached by Rtachyon, it must be:\begin{eqnarray*}
t^{-}<t^{+}+\frac{2d}{V_{t}^{-}} & \Longleftrightarrow & \frac{\bar{x}}{d}<\beta_{1}\frac{1-\beta_{t}\beta}{\beta_{t}-\beta}.\end{eqnarray*}

To let the measurements be uncorrelated both (a) and (b) must be true,
that is both measurements must take place before the arrival of the
correlating tachyon which departed when the other measurement was
performed.

Setting $\Delta\equiv\frac{\bar{x}}{d}$, we will have uncorrelated
measurements if

\[
-\beta_{1}\frac{1+\beta_{t}\beta}{\beta_{t}+\beta}<\Delta<\beta_{1}\frac{1-\beta_{t}\beta}{\beta_{t}-\beta}.\]

As it is well known, experiments testing Bell's inequality almost
always gave correlated measurements. This could be due to the fact
that the investigated $\Delta$ values were almost always outside
the range that gives uncorrelation (and this, as we'll see, could
be very likely). As far as I know, only in three cases \cite{b}-\cite{d}
changes on $\Delta$ value were experimentally investigated. Faraci
and others (1974) \cite{b} observed, in one of their five measurements,
a statistically relevant decrement of the correlation value. We will
analyze this point in more details in the part 2.\vspace {3mm}

Setting

\begin{equation}\Delta_{m}=-\beta_{1}\frac{1+\beta_{t}\beta}{\beta_{t}+\beta}\hspace{10mm};\hspace{10mm}\Delta_{M}=\beta_{1}\frac{1-\beta_{t}\beta}{\beta_{t}-\beta}\label{2}\end{equation}

it could be easily demonstrated that, in the assumed conditions, $0<\beta_{1}<1$,
$-1<\beta<1$ and $\beta_{t}>1$, the following relations occur:

\begin{eqnarray*}
\Delta_{M}>\Delta_{m}, & -1<\Delta_{m}<1, & -1<\Delta_{M}<1.\end{eqnarray*}

Whatever the values of $\beta$, $\beta_{t}$ and $\beta_{1}$, it
will always be possible to find an interval of $\Delta$ values, included
between -1 and 1, for which uncorrelated measurements are obtained.
That is, by opportunely varying $\bar{x}$ inside the interval $\left(-d,d\right)$
it will always be possible to individuate an interval of values for
which measurements will be uncorrelated.

Figures 1, 2 and 3 schematically show the three cases $\bar{x}<\Delta_{m}d$,
$\Delta_{m}d<\bar{x}<\Delta_{M}d$ and $\bar{x}>\Delta_{M}d$, figure
4 shows the same three cases in the Minkowski space.

\includegraphics[%
  scale=0.25]{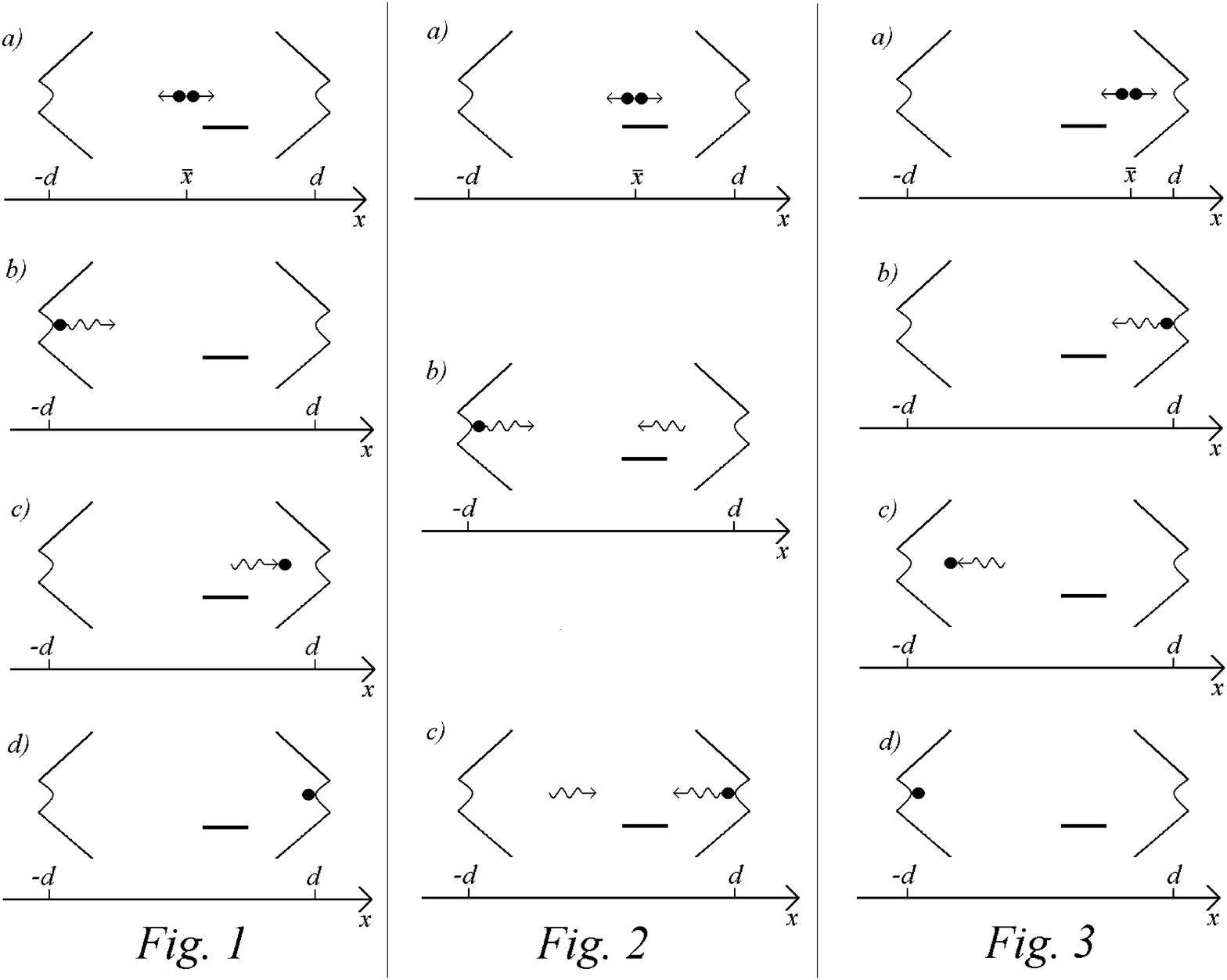}

\scriptsize

Figure 1: \emph{a) The couple of particles is created in $\bar{x}<\Delta_{m}d$;
b) Lparticle reaches the detector and Ltachyon leaves toward Rparticle;
c) Ltachyon reaches Rparticle; d) Rparticle reaches the detector after
the reception of Ltachyon. Measurements are correlated.}

\emph{Figure 2: a) The couple of particles is created in $\Delta_{m}d<\bar{x}<\Delta_{M}d$;
b) Rparticle reaches the detector before the reception of Ltachyon;
c) Lparticle reaches the detector before the reception of Rtachyon.
Measurements are} not \emph{correlated.}

\emph{Figure 3: a) The couple of particles is created in $\bar{x}>\Delta_{M}d$;
b) the Rparticle reaches the detector and Rtachyon leaves toward Lparticle;
c) Rtachyon reaches Lparticle; d) Lparticle reaches the detector after
the reception of Rtachyon. Measurements are correlated.}\normalsize\vspace{2mm}\\

\includegraphics[%
  scale=0.2]{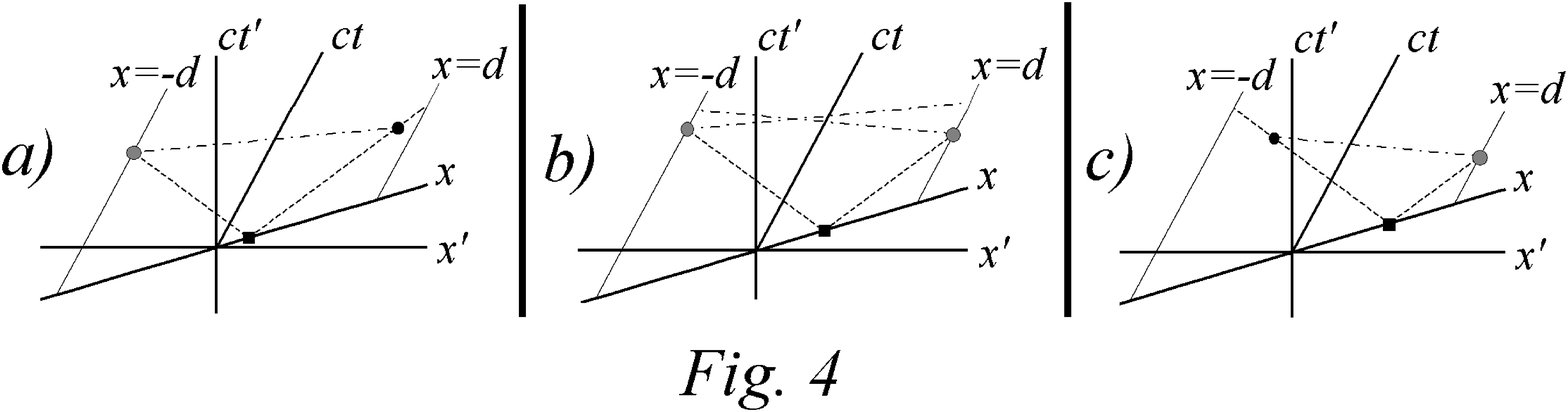}

\scriptsize

Figure 4: \emph{The curves was obtained by setting} $\beta_{1}=1$,
\emph{}$\beta_{t}=8$ \emph{and} $\beta=-0.4$ \emph{(we remember
that $\beta c$ is the velocity of} R' \emph{with respect to} R\emph{,
so the velocity of} R \emph{with respect} R' \emph{is} $-\beta c$\emph{).
By using eqs.} (\ref{2}) \emph{we get} $\Delta_{m}=11/38$$\simeq0.29$
\emph{and} $\Delta_{M}=0.5$\emph{.}

\emph{a) The couple of particles is created in $\bar{x}=0.2d<\Delta_{m}d$
(squared spot); Lparticle reaches the detector and Ltachyon leaves
toward Rparticle (gray circled spot); Ltachyon reaches Rparticle before
its detection in $x=d$ (black circled spot). We may notice that,
in R, Ltachyon is travelling {}``backward in time'' .Measurements
are correlated.}

\emph{b) The couple of particles is created in $\Delta_{m}d<\bar{x}=0.42d<\Delta_{M}d$
(squared spot); Rparticle reaches the detector before the reception
of Ltachyon and Lparticle reaches the detector before the reception
of Rtachyon (gray circled spots).Measurements are} not \emph{correlated.}

\emph{c) The couple of particles is created in $\bar{x}=0.6d>\Delta_{M}d$
(squared spot); Rparticle reaches the detector and Rtachyon leaves
toward Lparticle (gray circled spot); c) Rtachyon reaches Lparticle
before its detection in $x=-d$ (black circled spot). Measurements
are correlated.}\normalsize\vspace{3mm}\\

Experimentally detecting $\Delta_{m}$ and $\Delta_{M}$ it will be
possibile to obtain, for a given $\beta_{1}$, the values $\beta$
and $\beta_{t}$.

Setting $\beta_{1}=1$, that is assuming to perform the experiment
with a couple of photons, by the inversion of (\ref{2}) we obtain:\[
\beta_{t}^{2}-2\frac{1-\Delta_{M}\Delta_{m}}{\Delta_{M}-\Delta_{m}}\beta_{t}+1=0\]
\[
\left(\Delta_{M}+\Delta_{m}\right)\beta^{2}+2\left(1+\Delta_{M}\Delta_{m}\right)\beta+\left(\Delta_{M}+\Delta_{m}\right)=0.\]

The first equation gives the only acceptable solution\[
\beta_{t}=\frac{1-\Delta_{M}\Delta_{m}}{\Delta_{M}-\Delta_{m}}+\sqrt{\left(\frac{1-\Delta_{M}\Delta_{m}}{\Delta_{M}-\Delta_{m}}\right)^{2}-1}\]
 because the other solution, in the hypothesis $-1<\Delta_{m}<1$
and $-1<\Delta_{M}<1$, is less than 1 so not acceptable. If $\Delta_{M}\gtrapprox\Delta_{m}$
we obtain \begin{equation}
\beta_{t}\simeq2\frac{1-\Delta_{M}\Delta_{m}}{\Delta_{M}-\Delta_{m}}.\label{eq:9}\end{equation}

The second equation gives the only acceptable solution\begin{eqnarray*}
\beta=-\frac{1+\Delta_{M}\Delta_{m}}{\Delta_{M}+\Delta_{m}}+\sqrt{\left(\frac{1+\Delta_{M}\Delta_{m}}{\Delta_{M}+\Delta_{m}}\right)^{2}-1} & if & \Delta_{M}+\Delta_{m}>0,\end{eqnarray*}
\begin{eqnarray*}
\beta=0 & if & \Delta_{M}+\Delta_{m}=0,\end{eqnarray*}
\begin{eqnarray*}
\beta=-\frac{1+\Delta_{M}\Delta_{m}}{\Delta_{M}+\Delta_{m}}-\sqrt{\left(\frac{1+\Delta_{M}\Delta_{m}}{\Delta_{M}+\Delta_{m}}\right)^{2}-1} & if & \Delta_{M}+\Delta_{m}<0,\end{eqnarray*}
because the other solution, in the hypothesis $-1<\Delta_{m}<1$ and
$-1<\Delta_{M}<1$, is greater than 1 or less than -1 so not acceptable.
If $\Delta_{M}\gtrapprox\Delta_{m}$ (that is $\beta_{t}\gg1$) we
obtain \begin{equation}
\beta\simeq-\frac{\Delta_{M}+\Delta_{m}}{2}.\label{eq:10}\end{equation}

We notice that the interval of the acceptable $\Delta$ to obtain
uncorrelated measurements is given by $\Delta_{M}-\Delta_{m}=2\beta_{1}\beta_{t}\frac{1-\beta^{2}}{\beta_{t}^{2}-\beta^{2}}$
so, for all the acceptable $\beta_{1}$ and $\beta$ values, it will
always be

\[\lim_{\beta_{t}\rightarrow+\infty}\left(\Delta_{M}-\Delta_{m}\right)=0\  \] 

that is, for all kind of particles used in the experiment (photons
or not) and for any speed of \emph{R'} with respect to \emph{R}, the
interval of $\bar{x}$ values to obtain decorrelated measurements
will be very little (compared with \emph{d}) if the magnitude of the
speed of the tachyons in \emph{R'} is very big (compared with \emph{c})%
\footnote{This fact could be perhaps very interesting from a philosophical point
of view (if experiments always show correlations we could anyway believe
on local tachyons interactions that are not experimentally proven,
or not yet proven, because $\beta_{t}$ value is too high) but, of
coarse, would give a situation not very interesting from a physical
point of view.%
}.

We finally notice that, if things go the way hereby supposed, in the
experimental conditions in which uncorrelated measurements occur,
any conservation law will sure be violated. For example, measuring
spins of $e^{+}$ $e^{-}$ couples, it happens that, even if spin
detectors are placed along the same direction%
\footnote{To have uncorrelated measurements there will be no need to casually
change the direction of the detectors and this will probably give
a big simplification in the experimental set up compared with the
usual delayed-choice experiments.%
}, all the results, (+;+), (+;-), (-;+) and (-;-) will be possible
with the same probability%
\footnote{Or with a probability depending somehow on the $\Delta$ value. However,
if $\Delta_{m}<\Delta<\Delta_{M}$, probability of results (+;+) and
(-;-) will not be zero being this fact the experimental proof of the
uncorrelation of the measurements.%
}, giving an evident violation, in two of the four possible results,
of the angular momentum conservation law%
\footnote{We could anyway imagine that such violation disappears taking into
account angular momentum of experimental devices too.%
}. Parity should be obviously violated because the interval of $\bar{x}$
values giving uncorrelated measurements is symmetric with respect
to the origin only when $\beta=0$.\newpage

\section{Experimental situation.}

Let's go now to analyse experimental results \cite{b}-\cite{d}.

We must consider the tri-dimensional situation. We call $\overrightarrow{v}=\overrightarrow{\beta}c$
the velocity of the preferred tachyon reference frame \emph{R'} with
respect to the laboratory reference frame \emph{R} (see figure 5).

\includegraphics[%
  scale=0.2]{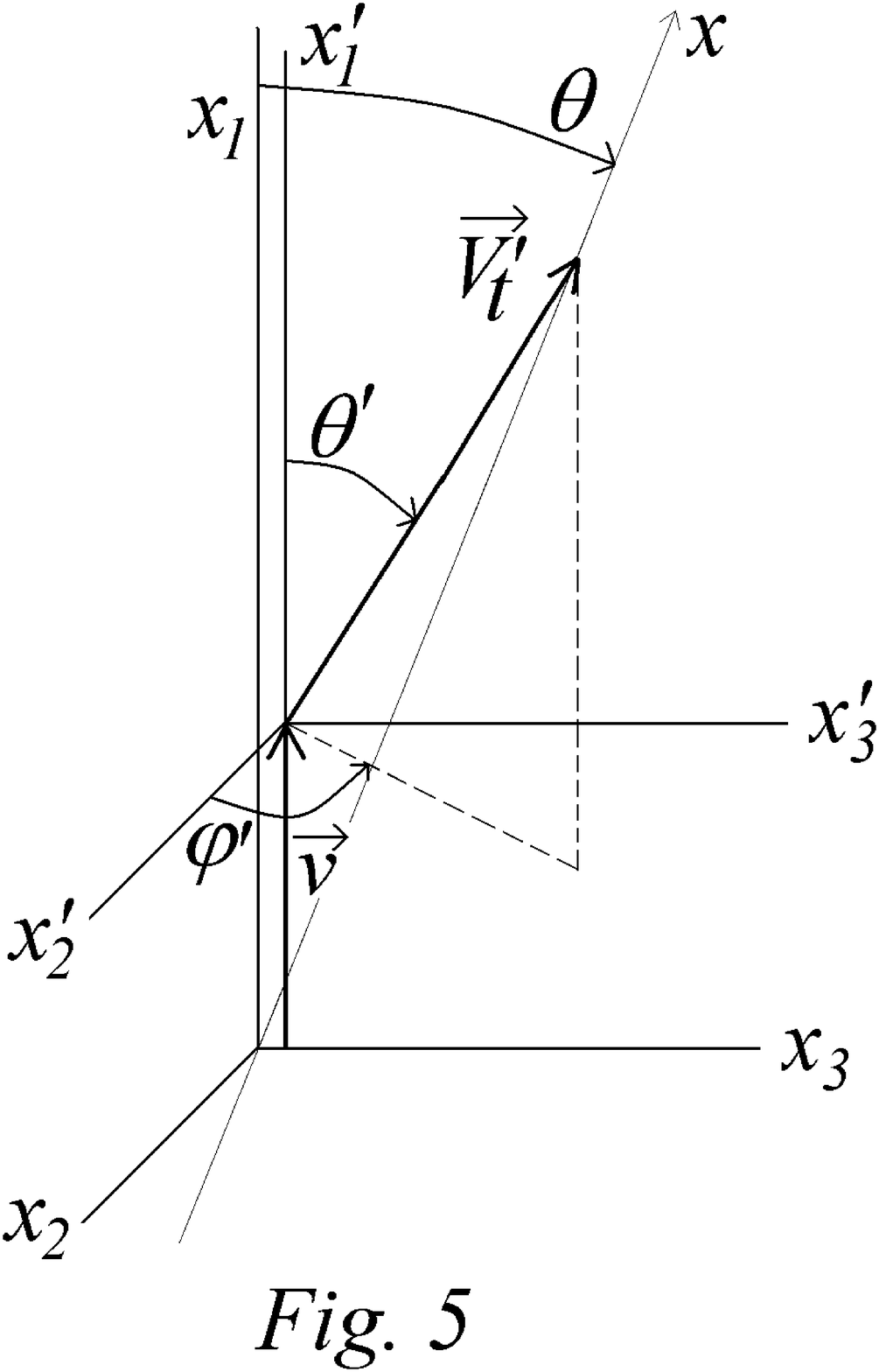}

Entangled particles flight is related to a certain \emph{x} axis in
the laboratory reference frame \emph{R}. The \emph{x} axis is defined
by $\theta$ and $\varphi$ angles, and we choose its orientation
in such a way that $0<\theta<\frac{\pi}{2}$. We want obtain the tachyon
velocity toward both the \emph{x} axis directions: $V_{t}^{+}$ and
$V_{t}^{-}$. $V_{t}^{+}$ is related to $\theta$ and $\varphi$
whilst $V_{t}^{-}$ is related to $\overline{\theta}=\pi-\theta$
and $\overline{\varphi}=\varphi+\pi$. Because of to the axial symmetry,
$\varphi$ does not enter the computation, so, following J. D. Jackson
(1975) \cite{h} , eq (11.32), we obtain:\begin{equation}
\tan\theta=\frac{\beta_{t}\sin\theta'}{\gamma\left(\beta_{t}\cos\theta'+\beta\right)}\label{eq:3}\end{equation}

and\begin{equation}
V_{t}^{+}=\frac{\sqrt{\beta_{t}^{2}+\beta^{2}+2\beta_{t}\beta\cos\theta'-\left(\beta_{t}\beta\right)^{2}\left(1-\cos^{2}\theta'\right)}}{1+\beta_{t}\beta\cos\theta'}c\label{eq:4}\end{equation}
where we setted $\gamma=\frac{1}{\sqrt{1-\beta^{2}}}$.

By inversion of (\ref{eq:3}) we may express $\cos\theta'$ in terms
of $\theta$ so, after substituting in (\ref{eq:4}), we finally get
the wanted relations.

We show the calculation results in the assumption $\beta_{t}\gg1$%
\footnote{In the reality we are assuming $\beta_{t}\cos\theta'\gg1$, that is,
in the limit $\beta_{t}\gg1$, our approximation is not good only
for $\theta'\simeq\frac{\pi}{2}$. The angles $\theta'\simeq\frac{\pi}{2}$
are important in the case $\gamma\gg1$, so, in the following, we
are assuming that $\gamma\gg1$ does not occur. Of coarse only experimental
results could prove the validity of such assumption.%
}.

When $\tan\theta>0$, as in the case of $V_{t}^{+}$, the inversion
of (\ref{eq:3}) gives $\cos\theta'=\frac{1}{\sqrt{1+\gamma^{2}\tan^{2}\theta}}$,
and, substituting in (\ref{eq:4}), we get:\begin{equation}
V_{t}^{+}=\frac{\frac{\beta_{t}}{\left|\cos\theta\right|\sqrt{1+\gamma^{2}\tan^{2}\theta}}}{1+\frac{\beta_{t}\beta}{\sqrt{1+\gamma^{2}\tan^{2}\theta}}}c.\label{eq:5}\end{equation}
When $\tan\overline{\theta}<0$, as in the case of $V_{t}^{-}$, the
inversion of (\ref{eq:3}) gives $\cos\theta'=\frac{-1}{\sqrt{1+\gamma^{2}\tan^{2}\overline{\theta}}}$,
and, substituting in (\ref{eq:4}), we get:\begin{equation}
V_{t}^{-}=\frac{\frac{\beta_{t}}{\left|\cos\overline{\theta}\right|\sqrt{1+\gamma^{2}\tan^{2}\overline{\theta}}}}{1-\frac{\beta_{t}\beta}{\sqrt{1+\gamma^{2}\tan^{2}\overline{\theta}}}}c.\label{eq:6}\end{equation}
Remembering that we choose \emph{x} axis direction so as to give $\cos\theta>0$,
$\overline{\theta}=\pi-\theta$, and setting\begin{eqnarray}
\beta^{*}\equiv\beta\cos\theta, & \textrm{$\gamma^{*}\equiv\frac{1}{\sqrt{1-\left(\beta^{*}\right)^{2}}},$} & \beta_{t}^{*}\equiv\frac{\beta_{t}}{\cos\theta\sqrt{1+\gamma^{2}\tan^{2}\theta}}=\beta_{t}\frac{\gamma^{*}}{\gamma},\label{eq:7}\end{eqnarray}
we could rewrite (\ref{eq:5}) and (\ref{eq:6}) in the forms%
\footnote{Because of the approximation $\beta_{t}\gg1$ we could add or subtract
$\beta^{*}$ at the numerator of eqs (\ref{eq:8}).%
}:\begin{eqnarray}
V_{t}^{+}=\frac{\beta_{t}^{*}+\beta^{*}}{1+\beta_{t}^{*}\beta^{*}}c & , & V_{t}^{-}=\frac{\beta_{t}^{*}-\beta^{*}}{1-\beta_{t}^{*}\beta^{*}}c.\label{eq:8}\end{eqnarray}
Comparing (\ref{1}) with (\ref{eq:8}) we can finally conclude that
the one-dimensional problem is equivalent to the tri-dimensional one
upon substituting $\beta_{t}$ with $\beta_{t}^{*}$ and $\beta$
with $\beta^{*}$. So, given $\theta$, $\beta$ and $\beta_{t}\gg1$,
we have an uncorrelating region centered in $\overline{\Delta}\equiv\frac{\Delta_{M}+\Delta_{m}}{2}$
whose largeness is $d\Delta\equiv\Delta_{M}-\Delta_{m}$ (see eqs.
(\ref{eq:9}) and (\ref{eq:10})):\begin{eqnarray}
\overline{\Delta}\simeq-\beta^{*} & , & d\Delta\simeq\frac{2}{\beta_{t}^{*}\left(\gamma^{*}\right)^{2}}\label{eq:11}\end{eqnarray}
where $\beta^{*}$, $\gamma^{*}$ and $\beta_{t}^{*}$ are given by
(\ref{eq:7}).

Because of Earth motion the $\theta$ angle changes with a period
of approximatively 1 day. This motion change the $\overline{\Delta}$
value and, if $\beta_{t}\gg1$ (so $\beta_{t}^{*}\gg1$ and $d\Delta\ll1$),
it could be very difficoult to get an experimental situation in which
the particle source is inside the uncorrelating region%
\footnote{We could direct \emph{x} axes (that is particle flight direction)
parallel to the Earth axes, so $\theta$ angle should remain constant.
Anyway, if $\beta_{t}$ is too big then $d\Delta$ could be so little
to give no possibility to the experimental proof of the existence
of the uncorrelating region.%
}. Anyway we have no theoretical indication on the $\beta_{t}$ value,
so we could think that it is high but not so high to give no possibility
to the experimental proof. Moreover, we could read the already performed
experiments and, since in one of them \cite{b} a statistically relevant
decrement of the correlation factor was already observed, we could
hope that decrement was due to the tachyons effects here supposed.

Faraci and others (1974) \cite{b} gave the results of five measurements:
three of them related to the same $\Delta$ value (they changed only
\emph{d} leaving $\bar{x}=0$) the other two related to $\Delta\simeq0.37$
and $\Delta\simeq0.72$. The four results related to $\Delta\simeq0$
and $\Delta\simeq0.37$ gave the same correlation value but the fifth
one gave a clearly lower value%
\footnote{As far as I know standard quantum mechanics interpretation does not
give any explanation of this correlation value decrement. I want also
outline that Faraci experiment showed a very good repeteability in
the four results related to $\Delta\simeq0$ and $\Delta\simeq0.3$7.%
}. We will now analyze this fifth measurement by supposing that the
correlation decrement was due to tachyons effect. By comparing the
lower correlation value with the other four we could argue that, during
the fifth measurement, the source spent $1/3$ of the total time inside
the uncorrelating region. The total time for each measurement was
several weeks%
\footnote{G. Faraci: private communication.%
} so the $\theta$ variation due to Earth motion was well mediated.
We call $d\overline{\Delta}$ the daily variation of $\overline{\Delta}$,
and we saw that it should be three times the uncorrelating region
amplitude, $d\Delta$. Since $d\Delta$ is little%
\footnote{$\beta_{t}\gg1$ so $d\Delta\ll1$ and $\beta_{t}$ must be very much
greater than 1 because if $d\Delta$ was not very little then the
effects here conjectured should be already observed several times.%
} also $d\overline{\Delta}\simeq3d\Delta$ must be little. The variation
$d\overline{\Delta}$ could be little only if the direction of $\overrightarrow{v}$
(the velocity of the tachion aether with respect to the laboratory
reference frame) is close to the Earth axis. We call $\delta$ the
angle between the direction of $\overrightarrow{v}$ and the Earth
axis and we suppose $\delta\ll1$. The experiment took place on Catania
(Italy) and \emph{x} axes (the direction of the entangled particle
flight path) was close to the South-North direction%
\footnote{G. Faraci: private communication. I thank very much G. Faraci for
these communications.%
}. Taking into account the Catania latitude value $\theta_{C}=37^{o}30'$,
we could say that because of Earth motion $\theta$ changed from $\theta_{C}-\delta$
to $\theta_{C}+\delta$ (we assume that the direction of the entangled
particle flight path was exactly the South-North). So the variation
$d\overline{\Delta}$ was approximatively given by (see (\ref{eq:7})
and (\ref{eq:11})):\[
d\overline{\Delta}\simeq2\delta\beta\sin\theta_{C}.\]
 By imposing that $\Delta\simeq0.72$, the experimental value of $\Delta$
that gave less correlation, must be inside the interval $\overline{\Delta}\pm\frac{d\overline{\Delta}}{2}$
we get:\[
\beta=\frac{0.72}{\cos\theta_{C}}\pm\delta\tan\theta_{C}=0.91\pm0.76\delta.\]
Unfortunately we have no indication on the $\delta$ value. We just
know, as already noticed, that it should be {}``little''. Only the
repetition of the experiment to directly measure the $d\overline{\Delta}$
amplitude could give a realable $\delta$ value. Once known $\delta$,
we get $\beta_{t}$ value by setting $d\overline{\Delta}\simeq3d\Delta$
(see see (\ref{eq:7}) and (\ref{eq:11})):

\[
\beta_{t}\simeq\frac{3\gamma}{\delta\beta\sin\theta_{C}\left(\gamma^{*}\right)^{3}}.\]

Of coarse, if the effects is real, it could be better revealed by
means of similar experiments that use faster statistics (minutes,
not weeks) so as, for example, A. Aspect and others (1981) \cite{f}.

As far as Wilson and others \cite{c} and Bruno and others \cite{d}
experiments are concerned I must notice that a realible comparison
is possible only after knowing photons flight direction (and I do
not know it). Since they never observed a decrement of correlation
value the only conclusion we may obtain by their results is that,
if Faraci and others results \cite{b} is due to tachyons hereby supposed,
then the daily variation of $\overline{\Delta}$, $d\overline{\Delta}$,
should be {}``little'' and, as a consequence, $\delta$ angle should
be {}``little'' too.\newpage

\section*{Appendix I}

Probably the first one to talk about causal paradoxes due to tachyons
was A. Einstein (1907) \cite{e}. Similar statements were later given
by W. Pauli in {}``Relativitätstheorie'' chapter 1.6, Leipzig (1921).
Both of them essentially deny tachyons possibility because speed composition
law already shown in this paper, $V_{t}^{+}=\frac{\beta_{t}+\beta}{1+\beta\beta_{t}}c$
and $V_{t}^{-}=\frac{\beta_{t}-\beta}{1-\beta\beta_{t}}c$, would
give signals propagating {}``backwards in time'' for suitable $\beta$
values. For example, if $\beta<-\frac{1}{\beta_{t}}$, we get $V_{t}^{+}<0$
and a signal which left from position $x=0$ at the instant $t_{in}$
(i. e. when the clock fixed in $x=0$ signs the instant $t_{in}$)
would reach point $x=d>0$ at the instant $t_{fin}=t_{in}+\frac{d}{V_{t}^{+}}<t_{in}$
(i. e. when the clock fixed in $x=d$ signs the instant $t_{fin}$).

In this form the statement is surely wrong. Conventionality of simultaneity
gives no physical meaning to the comparison of instants signed by
clocks fixed on different points%
\footnote{Or, as already noticed, if we would give them any meaning, we must
take into account the synchronization chosen, and the simple fact
that $t_{fin}<t_{in}$ does not give any causal paradox even after
taking into account the standard synchronization assumed by Einstein
and Pauli.%
}.

Later the discussions on tachyons and causal paradoxes assumed a little
bit more subtle form (I am not able to say if this form was implicitly
supposed by Einstein and Pauli). Substantially, we assume that at
least one reference frame in which tachyons propagate isotropically
exists, then, \emph{assuming that the relativity principle is suitable
for tachyons}, the fact that tachyons could propagate isotropically
is extended to all inertial reference frames. Moreover, still because
of relativity principle, it can be shown that if in a certain reference
frame tachyons could propagate toward the \emph{positive} \emph{x}
axes direction at a certain velocity, then, \emph{in any other reference
frame}, they could propagate toward the \emph{negative} \emph{x} axes
direction at the same velocity.

This gives the following paradox:

the frame $R_{2}$ moves, respect to $R_{1}$, at the $\beta c$ ($-1<\beta<1$)
velocity toward the positive direction of the \emph{x} axes. Let's
call $\beta^{*}c$ ($\beta^{*}>1$) the tachyon velocity toward the
positive \emph{x} axes direction in $R_{1}$. Because of what we just
saw, in $R_{2}$ tachyons could propagate at a $\beta^{*}c$ velocity
toward the negative direction of the \emph{x} axes. From composition
velocity law it follows that tachyons propagating in $R_{2}$ at a
$\beta^{*}c$ velocity toward the negative \emph{x} axes, have in
$R_{1}$ a velocity (in the negative \emph{x} direction) given by
$\frac{\beta^{*}-\beta}{1-\beta^{*}\beta}c$. This means that, if
in $R_{1}$ we send a going and returning tachyon along the segment
\emph{AB} of length \emph{d} (we suppose also here that clocks were
synchronized by standard relation) we'll obtain that, if the going
tachyon, the one travelling at velocity $\beta^{*}c$, leaves \emph{A}
when the clock fixed in \emph{A} was signing the instant $t_{in}$,
then it will arrive in \emph{B} when the clock fixed in \emph{B} signs
the instant $t_{in}+\frac{d}{\beta^{*}c}$; moreover, if the returning
tachyon, the one travelling at velocity $\frac{\beta^{*}-\beta}{1-\beta^{*}\beta}c$,
leaves \emph{B} when the clock fixed in \emph{B} was signing the instant
$t_{in}+\frac{d}{\beta^{*}c}$ (that is the returning tachyon leaves
from \emph{B} simultaneously with the arrival in \emph{B} of the going
tachyon), then it will arrive in \emph{A} when the clock fixed in
\emph{A} signs the instant $t_{fin}=t_{in}+\frac{d}{\beta^{*}c}+\frac{1-\beta^{*}\beta}{\beta^{*}-\beta}\frac{d}{c}$.

By easy calculations we obtain $t_{fin}-t_{in}=\frac{d}{\beta^{*}c}\frac{2\beta^{*}-\beta\left(1+\beta^{*}\right)}{\beta^{*}-\beta}$
that is positive for every $\beta$ ($-1<\beta<1$) if $\beta^{*}<1$
whilst it is negative for $\beta>\frac{2\beta^{*}}{1+\left(\beta^{*}\right)^{2}}$
if $\beta^{*}>1$ as assumed here. Being $\frac{2\beta^{*}}{1+\left(\beta^{*}\right)^{2}}<1$
for every $\beta^{*}$ value, we obtain that, for each velocity of
the going tachyon in $R_{1}$, it will always be possible to find
a frame $R_{2}$ (moving at velocity $\beta c$, with $0<\beta<1$,
with respect to $R_{1}$) from which we could send a returning tachyon
that will reach \emph{A} before the departure of the going tachyon.

The only chance to avoid the causal paradox seems to be the choice
$\beta^{*}<1$ that makes $t_{fin}-t_{in}$ always positive. That
is, the only choice to avoid causal paradoxes seems to be the deny
of tachyons possibility.

It is clear that we cannot invoke here the simultaneity conventionality
because the comparison $t_{fin}-t_{in}$ does not concern now two
different clocks, but the same clock: the clock fixed in \emph{A}.

It is also clear that the demonstration shown above is hardly based
on relativity principle. It is enough to say that such principle is
not suitable for tachyons (we could for example imagine that, as far
as tachyons are concerned, we are not {}``sotto coverta'') and the
shown demonstration on causal paradoxes is no longer valid. Sure the
paradox disappeares if, as it is here supposed, we imagine that tachyons
have a preferred frame, an {}``aether'' in which they propagate
isotropically. And, in the frames moving respect to the aether, tachyons'
velocity will be given by the velocity composition law. In such a
case the tachyon back and forth travel time on a segment of length
\emph{d}, $\frac{d}{V_{t}^{+}}+\frac{d}{V_{t}^{-}}$, is positive
for each $\beta_{t}>0$, i. e. no causal paradox will be given by
tachyons existence in the hypothesis that they propagate in their
{}``aether''.

\section*{Appendix II}

In this appendix my intent is not to demonstrate that tachyons \emph{must}
be related to a certain momentum. I want only to explain why, in my
opinion, if tachyons do exist, then a suitable momentum should be
related to them. Moreover, I also want to show which could be a suitable
tachyon momentum definition.

My opinion is that if we want to give a physical meaning to the propagation
of any particle (as well as any signal) we must say which is the measurement
to perform to know the direction of such propagation (that is which
is the physical entity that describe the signal propagation direction).
And the propagation direction of any particle has surely a physical
meaning: it is the direction {}``from the cause toward the effect''.

Anderson and others \cite{a} clearly show that choosing a suitable
value of the synchronization 3-vector $\overrightarrow{\kappa}$ (\cite{a}
pag 127) we could set velocity%
\footnote{we are meaning {}``one-way'' velocity.%
} of any particle (mass particles as well as photons) to any value.
In my opinion this is enough to say that velocity is not a physical
meaning entity. So, for example, velocity does not give us the propagation
direction of a particle: it can happen sometimes%
\footnote{For example this happens if we sinchronize clocks by standard relation
(that is imposing the isotropy of the one-way speed of light) and
we never consider a signal faster than light.%
} but we cannot be sure that this always happens. So, if we think that
propagation direction of any signal has a physical meaning, we must
link the propagation direction of a signal to any other physical meaning
entity that is not its velocity. And this physical meaning entity
(as well as any other physical meaning entity), in my opinion, must
be not conventional because we cannot imagine to conventionally choose
a signal direction propagation, we cannot conventionally choose the
direction from the cause toward the effect.

Anderson and others (\cite{a} pag 127) show that synchronization
transformations affect the time component of any 4-vectors but do
not affect spatial components%
\footnote{I remember that I'm calling 4-vector the vector written by means of
its controvariant components.%
} as well as 4-vectors magnitude. In my opinion the synchrony invariance
has a deep physical meaning: the synchrony invariant entities are
measurable, so physical meaning entities, whilst not synchrony invariant
entities are not measurable%
\footnote{I do not intend to demonstrate here this. I am simply thinking on
it as a conjecture. Anyway, after \cite{a}, seems to me absolutely
clear that, who wants to reject simultaneity conventionality thesis,
must show the way to measure any not invariant under synchronization
transformations entity. Since I cannot imagine any measurement not
reducible to a lenght measurement, or to any scalar measurement, or
to a time interval measurement performed by a travelling clock, and
since all these entities are synchrony invariants, I cannot imagine
how simultaneity conventionality thesis could be rejected. That is
I cannot imagine how any physics law could be violated because of
changing synchronization or, equivalently, how any physics law may
obly us to choose a certain synchronization. If we change synchronization,
each physics law must be rewritten by means of \cite{a} pag 127 prescriptions,
and if a law is experimentally proven (by lenght, or scalar, or travelling
clock time interval measurements) in standard form it will be automatically
proven in non standard form too.%
}.

For example, as far as $X=\left(cdt,d\overrightarrow{x}\right)$ is
concerned, we have spacial components, $d\overrightarrow{x}$, and
magnitude%
\footnote{We assume standard synchronization.%
} $\left\Vert X\right\Vert ^{2}=\left(cdt\right)^{2}-\left|d\overrightarrow{x}\right|^{2}=\left(cd\tau\right)^{2}$
that are measurable entities: $d\overrightarrow{x}$ is measurable
by using a rigid rod, $d\tau$ is measurable by using a (uniformely)
travelling clock%
\footnote{$d\tau$ is the interval time measured by the travelling clock that
leaves the point $\overrightarrow{x}{}_{in}$ when the clock fixed
in $\overrightarrow{x}{}_{in}$ was signing the instant $t_{in}$
and arrives in the point $\overrightarrow{x}{}_{in}+d\overrightarrow{x}$
when the clock fixed in the point $\overrightarrow{x}{}_{in}+d\overrightarrow{x}$
is signing the instant $t_{in}+dt$. In transport synchronization
we use exactly such $d\tau$ measurement to fix clocks instants.%
}. As far as a mass particle 4-vector momentum is concerned, $P_{m}=\left(mc\gamma,\overrightarrow{p}\right)$,
both the magnitude $\left\Vert P_{m}\right\Vert ^{2}=\left(mc\right)^{2}$
and the spacial components $\overrightarrow{p}=m\frac{d\overrightarrow{x}}{d\tau}$
are measurable entities. As far as far a photon 4-vector momentum
is concerned, $P_{ph}=\left(h\frac{\omega}{c},h\overrightarrow{k}\right)$,
the magnitude $\left\Vert P_{m}\right\Vert =0$ by definition (that
is, in standard synchronization, we set clocks in such a way to have
$\left\Vert P_{m}\right\Vert =0$) and the spatial part, $\overrightarrow{k}$,
is measurable by means of interferometric experiments, that is by
means of length measurements.

We may notice that the spatial part of 4-vector momentum (in both
cases, mass particles so as photons) is a 3-vector having the direction
of the particle propagation. So, at least for mass particles and photons,
this is the physical entity we were looking for. If we want to use
only not conventional entities (that is if we want to express physical
meaning statements whatever the synchronization chosen) we must say
that mass particles, so as photons, propagate toward the direction
of their 3-vector momentum, not toward the direction of their velocity.
Changing synchronization we may change the velocity but we cannot
change the momentum, that is we cannot change the signals propagation
direction, that is we cannot change the direction from the cause toward
the effect. Of course we cannot. This direction is not conventional,
it is a matter of fact.

So, if tachyons do exist, and if their propagation have any physical
meaning, I imagine that also for tachyons it should be possible to
define a 4-vector momentum, and I imagine that the spacial components
of the tachyons 4-vector momentum give their propagation direction.

Since there not exists any inertial reference frame where a tachyon
can be at rest, so as for photons, I suppose that 4-vector momentum
definition is similar for tachyons and photons.

Let's imagine to be at rest with respect to the tachyons preferred
reference frame \emph{R'} where clocks was synchronized by standard
relations. We have that photon 4-vector momentum is given by

\[
P'_{ph}=\left(h\frac{\omega'_{ph}}{c},h\overrightarrow{k}'_{ph}\right)=\left(p'_{0,ph},\overrightarrow{p}'{}_{ph}\right)\]
and we suppose that tachyon 4-vector momentum $P'_{t}$ is given by\[
P'_{t}=\left(\bar{h}\frac{\omega'_{t}}{c},\bar{h}\overrightarrow{k}'_{t}\right)=\left(p'_{0,t},\overrightarrow{p}'_{t}\right)\]
where $\bar{h}$ is a suitable constant to be experimentally determined.

For mass particles, as well as as for photons, we can define velocity
by means of the ratio between spacial components and time component
of the 4-vector momentum. For example, for a photon we have $\overrightarrow{v}'_{ph}=c\frac{\overrightarrow{p}'_{ph}}{p'_{0,ph}}$
and we must transform $P'_{ph}$ into $P_{0,ph}=\left(p_{0,ph},\overrightarrow{p}{}_{ph}\right)$
by means Lorentz transformations to have the velocity in the laboratory
reference frame \emph{R}: $\overrightarrow{v}{}_{ph}=c\frac{\overrightarrow{p}{}_{ph}}{p{}_{0,ph}}$.
We can observe that $\overrightarrow{v}{}_{ph}$ and $\overrightarrow{p}{}_{ph}$
have the same direction or the opposte one depending on the sign of
$p_{0,ph}$. In standard synchronization this sign is always positive%
\footnote{$p_{0,ph}=\gamma\left(p'_{0,ph}+\overrightarrow{\beta}\cdot\overrightarrow{p}'_{ph}\right)=\gamma p'_{0,ph}\left(1+\overrightarrow{\beta}\cdot\frac{\overrightarrow{p}'_{ph}}{p'{}_{0,ph}}\right)$
where $\overrightarrow{\beta}c$ is the velocity of \emph{R'} with
respect to \emph{R}. Noticing that $\left|\overrightarrow{\beta}\right|<1$
and that, in standard synchronizaton, $\left|\overrightarrow{p}'_{ph}\right|=p'_{0,ph}$,
we obtain that if $p'_{0,ph}>0$ also $p_{0,ph}>0$ for any acceptable
$\overrightarrow{\beta}$.%
}, so we can conclude that, even if velocity is a conventional entity,
taking into account the synchronization chosen (the standard one),
we can give it a physical meaning (at least for light velocity): for
example we can say that its direction is always the same direction
of momentum. We can reach similar conclusion also for mass particles:
in standard synchronization their velocity has the same direction
of their momentum.

But for the tachyons here supposed (that is tachyons propagating in
their \emph{aether}) the situation is very different. In \emph{R'}
we know that, for any tachyon propagation direction, it is always
$\left|\overrightarrow{v}{'}_{t}\right|=\beta_{t}c$, so, if we define
tachyons velocity as well as for photons and mass particles\[
\overrightarrow{v}'_{t}=c\frac{\overrightarrow{p}'_{t}}{p'_{0,t}}\]
we obtain $p'_{0,t}=\frac{1}{\beta_{t}}\left|\overrightarrow{p}'_{t}\right|=\frac{1}{\beta_{t}}\bar{h}\left|\overrightarrow{k}'_{t}\right|$,
and we could rewrite $P_{t}$ in such a way:\[
P'_{t}=\left(\bar{h}\frac{\omega'_{t}}{c},\bar{h}\overrightarrow{k}'_{t}\right)=\bar{h}\left|\overrightarrow{k}'_{t}\right|\left(\frac{1}{\beta_{t}},\hat{n'}\right)\]
where $\hat{n'}$ is the versor directed toward the propagation direction.
By means of Lorentz transformations we can now obtain the tachyon
momentum 4-vector in the laboratory refence frame \emph{R}. We easily
see that the time component of such 4-vector, $p_{0,t}=\gamma\left(p'_{0,t}+\overrightarrow{\beta}\cdot\overrightarrow{p}'_{t}\right)=\gamma\bar{h}\left|\overrightarrow{k}'_{t}\right|\frac{1}{\beta_{t}}\left(1+\beta_{t}\overrightarrow{\beta}\cdot\hat{n'}\right)$,
can be negative for suitable $\overrightarrow{\beta}$ ($\left|\overrightarrow{\beta}\right|<1$),
that is there exist inertial reference frames where $\overrightarrow{v}{}_{t}=c\frac{\overrightarrow{p}{}_{t}}{p{}_{0,t}}$
and $\overrightarrow{p}{}_{t}$ have opposite directions. In such
cases, if we synchronize clocks by standard relation, tachyons propagate
in a certain direction (the $\overrightarrow{p}{}_{t}$ direction)
even if they {}``are seen'' propagating {}``backward in time'',
as well as the plane travelling from New Zealand to Italy.

\vspace {8mm}

\textbf{\emph{Acknowledgements}}\emph{: I cannot find suitable words
to express my thankfulness to E. Fabri for all the help he gave me
in every physics argument I tried to understand by discussing on Usenet
forums.}

\emph{Public Usenet discussion with \char`\"{}El filibustero\char`\"{}
was for me important to understand the common vision on causal paradoxes
due to tachyons. I am very grateful to him.}

\emph{Thank you to A. Cintio, L. Fronzoni, V. Moretti as well as to
my friend and teacher S. Faetti for the useful remarks they gave me
after reading the preprint of this paper.}

\emph{Thank you very much to my friend A. Rossi for the help with
the English language.}

\emph{Finally a special thank for the help with the figures and for
very much other to my daugther Irene.}   

\newpage

\begin{thebibliography}{99}\bibitem{f} A. Aspect, P. Grangier and G. Roger\\ \textsl{Experimental tests of realistic local theories via Bell's theorem}\\ Phys. Rev. Lett. \textbf{47}, 460 (1981).\bibitem{i} G. C. Ghirardi \\ \textsl{Un'occhiata alle carte di Dio}\\ Il Saggiatore, Milano (1997).\bibitem{g} F. Selleri \\ \textsl{La fisica del Novecento. Per un bilancio critico}\\ Progedit, Bari (1999).\bibitem{a} R. Anderson, I. Vetharaniam, G. E. Stedman\\ \textsl{Conventionality of synchronisation, gauge dependence and test theories of relativity}\\ Phys. Rep. \textbf{295}, 93-180 (1998). \bibitem{b} G. Faraci, D. Gutkowski, S. Notarrigo and A. R. Pennisi \\ \textsl{An Experimental Test of the EPR Paradox}\\ Lettere al Nuovo Cimento \textbf{9}, 15, 607-611 (1974). \bibitem{c} A R Wilson, J Lowe and D K Butt \\ \textsl{Measurement of the relative planes of polarization of annihilation quanta as a function of separation distance}\\ J. Phys. G: Nucl. Phys. \textbf{2}, 613-624 (1976).\bibitem{d} M. Bruno, M. D'Agostino and C. Maroni \\ \textsl{Measurement of Linear Polarization of Positron Annihilation Photons}\\ Nuovo Cimento \textbf{40B}, 143 (1977). \bibitem{e}  A. Einstein \\ \textsl{Über die vom Relativitätsprinzip geforderte Trägheit der Energie}\\ Annalen der Physik \textbf{23}, 371-384 (1907).\bibitem{h} John D. Jackson \\ \textsl{Classical Electrodynamics}\\ John Wiley and Sons, Inc. (1975) .\end{thebibliography}
\end{document}